\documentclass[aps,pra,eqsecnum,twocolumn]{revtex4}
\usepackage{graphicx}
\usepackage{bm}
\begin{document}
\preprint{}
\title{Nonclassicality of states and measurements
by breaking classical bounds on statistics}
\author{\'{A}ngel Rivas}
\email{A.Rivas@herts.ac.uk}
\affiliation{School of Physics, Astronomy and Mathematics,
University of Hertfordshire College Lane, Hatfield, Hertfordshire,
AL10 9AB, United Kingdom}
\author{Alfredo Luis}
\email{alluis@fis.ucm.es}
\homepage{http://www.ucm.es/info/gioq}
\affiliation{Departamento de \'{O}ptica, Facultad de Ciencias
F\'{\i}sicas, Universidad Complutense, 28040 Madrid, Spain }
\date{\today}

\begin{abstract}
We derive exceedingly simple practical procedures revealing the
quantum nature of states and measurements by the violation of
classical upper bounds on the statistics of arbitrary measurements.
Data analysis is minimum and definite conclusions are obtained
without evaluation of moments, or any other more sophisticated
procedures. These nonclassical tests are independent of other
typical quantum signatures such as sub-Poissonian statistics,
quadrature squeezing, or oscillatory statistics. This approach
can be equally well applied to very diverse situations such as
single- and two-mode fields, observables with continuous and
discrete spectra, finite- and infinite-dimensional systems, and ideal
and noisy measurements.
\end{abstract}


\maketitle

\section{Introduction}

Nonclassicality is a key concept supporting the necessity of the
quantum theory \cite{MSZ,MH,ase,as,AT,vyt,SP,MM}. A customary
signature of nonclassical behavior is the failure of the
Glauber-Sudarshan $P$ phase-space representation to exhibit all
the properties of a classical probability density. This occurs
when $P$ takes negative values, or when it fails to be a proper
function becoming a generalized function or distribution.

Within standard quantum theory, quantum states play two dissimilar
but complementary roles: (i) they express the state of the system,
and (ii) they determine the statistics of measurements by projection
on the system state, such as, for example, photon-number and quadrature
measurements in quantum optics. We may refer to them as measured and
measuring states respectively.

In this work we derive exceedingly simple and robust practical
procedures to reveal the quantum nature of measured and
measuring states. In this regard, while characterization of
nonclassical (measured) states has been well developed
\cite{MSZ,MH,ase,as,AT,vyt}, much less attention has received the
characterization of measurements \cite{MM}. One of
the purposes of this work is to contribute to fill this gap addressing
the characterization of nonclassical measurements, i. e., when the
measuring state is nonclassical. More specifically, measurements
are described by positive operator-valued measures (POVMs) $\Delta_m$,
such that the statistics of the measurement is $p_m = \textrm{tr} (
\Delta_m \rho )$, where $\rho$ is the measured state. We will say
that the measurement is nonclassical when the $P$ representative of
some $\Delta_m$ takes negative values or is a generalized function.
In most practical situations, $\Delta_m$ define legitimate measuring
states $\rho_m \propto \Delta_m$ so that the measurement is
nonclassical if and only if some $\rho_m$ is nonclassical.
Nonclassicality of measurements has been recently related with the
noncontextuality problem in Ref. \cite{SP}.

The main contributions of this work are as follows:

(i) We derive exceedingly simple practical procedures that can
reveal the quantum nature of states and measurements. These are
upper bounds on measurement statistics which are satisfied by
all states and measurements for which the $P$ representative is
a non-negative function compatible with classical physics. The
lack of compliance of these statistical bounds is thus a
nonclassical signature.

(ii) This approach can be applied to arbitrary measurements, which
may involve for example single- or two-mode electromagnetic fields,
observables with continuous or discrete spectrum, systems on finite-
or infinite-dimensional spaces, ideal or noisy measurements, etc.
(Some of these possibilities are considered in detail below.) This is
in sharp contrast with other nonclassical criteria that refer
exclusively to specific measuring schemes.

(iii) A key point of this approach is that data analysis is
reduced to minimum. At difference with other tests of nonclassical
behavior, in our case definite conclusions can be obtained without
evaluation of moments, or any other more sophisticated data
elaborations \cite{MSZ,MH,ase,as,AT,vyt}. This is reflected on the
robustness under practical imperfections that may even favour
observation of nonclassical behavior.

(iv) These nonclassical tests are independent of other typical
quantum signatures of nonclassical behavior such as sub-Poissonian
statistics, squeezing, or oscillatory statistics \cite{MSZ}. To
this end we propose examples of quantum states violating classical
bounds that present no such typical quantum signatures.

To derive the nonclassical tests we will use the $P$ and $Q$
phase-space representatives associated to any operator $A$,
defined as
\begin{equation}
A = \int d^2 \alpha P(\alpha ) | \alpha \rangle \langle \alpha |,
\quad
Q(\alpha ) = \frac{1}{\pi} \langle \alpha | A | \alpha \rangle ,
\end{equation}
where $| \alpha \rangle$ are coherent states, $a |\alpha \rangle =
\alpha | \alpha \rangle$, and $a$ is the annihilation or
complex-amplitude operator. They are suitably normalized
\begin{equation}
\label{nor}
\int d^2 \alpha P (\alpha) = \int d^2 \alpha Q(\alpha)
= \textrm{tr} A,
\end{equation}
with $d^2 \alpha = d x d y $, where $x$, and $y$ are the real and
imaginary parts of $\alpha = x+iy$. The measured statistics $p_m
= \textrm{tr} ( \Delta_m \rho )$ can be then expressed as
\begin{equation}
\label{pm}
p_m = \pi \int d^2 \alpha P_m (\alpha) Q (\alpha) = \pi \int d^2
\alpha P (\alpha) Q_m (\alpha)  ,
\end{equation}
where $P (\alpha)$ and $Q (\alpha)$ are the $P$ and $Q$
representatives of the measured state $\rho$, while $P_m (\alpha)$
and $Q_m (\alpha)$ are the ones associated to the POVM $\Delta_m$.

In Secs. II and III we derive simple bounds to $p_m$ able to reveal
the nonclassical nature of measuring states $\rho_m \propto
\Delta_m$ and measured states $\rho$, respectively. The robustness
of these criteria under practical imperfections is examined in
Sec. IV. This formalism is further extended to two-mode situations
in Sec. V, and adapted to finite-dimensional systems in Sec. VI.

\section{Nonclassical measurements}

From Eq. (\ref{pm}) we can derive classical bounds disclosing
nonclassical measurements. For every ordinary non-negative
function $P_m (\alpha) \geq 0$ it holds that for every $\alpha$
\begin{equation}
\label{pPm}
P_m (\alpha) Q(\alpha) \leq P_m (\alpha) Q_{\mathrm{max}},
\end{equation}
where $Q_{\mathrm{max}}$ is the maximum of $Q(\alpha)$ (note
that $Q(\alpha)$ is always a positive and well behaved
function). Applying this to the first equality in Eq. (\ref{pm})
we get the following upper bound for $p_m$, provided that
$\textrm{tr} \Delta_m$ is finite,
\begin{equation}
\label{trPQ}
p_m  \leq \pi Q_{\mathrm{max}} \textrm{tr} \Delta_m .
\end{equation}
Equation (\ref{trPQ}) can be violated if $P_m (\alpha)$ fails
to be positive or when it becomes a generalized function. In
both cases Eq. (\ref{pPm}) fails to be true. Therefore, the
violation of condition (\ref{trPQ}) is a signature of
nonclassical measurement.

The existence of $P_m(\alpha)$ as a classical probability density
for all $m$ allows us to understand the measurement as
a classical stochastic process \cite{CM77} between the phase space
and the sample space, with transition probability kernel given by
$K(m,\alpha)= \pi P_m(\alpha)$. Conversely, the failure of
$K(m,\alpha)$ to be a classical conditional probability density
denotes the quantum nature of the measurement process.

In order to detect the violation of the classical bound
(\ref{trPQ}) the only prior information required about the
measurement being performed is the trace $\mathrm{tr}(\Delta_m)$.
This can be measured using explicit practical methods (see
some proposals in the Appendix). In any case this is
not a very stringent condition since in most practical situations
this can be inferred from simple rough analyses of the
experimental arrangement, by symmetry considerations, etc.

Note that coherent states $| \alpha \rangle$ are useless as
measured states to reveal nonclassical measurements since
$\pi Q_{\mathrm{max}} = 1$ so that  Eq. (\ref{trPQ}) leads to
the trivial bound $p_m \leq \textrm{tr} \Delta_m$ for all
measurements \cite{LL}. This bound is trivial because, using
the Cauchy-Schwarz inequality,
\begin{equation}
\left | \textrm{tr} (A B^\dagger) \right |^2
\leq \textrm{tr} (A A^\dagger ) \textrm{tr} (B B^\dagger ),
\end{equation}
we get
\begin{equation}
p_m^2 = \left [ \textrm{tr} \left ( \rho \Delta_m \right )
\right ]^2 \leq  \textrm{tr} \left ( \rho^2 \right )\textrm{tr}
\left ( \Delta^2_m \right ),
\end{equation}
and using that for positive operators $\textrm{tr} (A^2) \leq
( \textrm{tr} A )^2$ we get
\begin{equation}
p_m = \textrm{tr} (\rho \Delta_m ) \leq \textrm{tr} \rho  \;
\textrm{tr} \Delta_m = \textrm{tr} \Delta_m.
\end{equation}
Otherwise, quantum or classical state other than coherent
may be used since the weight of the criteria relies on the
behavior of $P_m (\alpha)$.

This approach is next illustrated with the examples of photon-number and field quadrature measurements performed on a single-mode electromagnetic field.

\subsection{Photon-number measurements}

In order to illustrate this formalism the simplest example is the
ideal photon-number measurement, $\Delta_n = \rho_n = | n \rangle
\langle n |$, where $| n \rangle$ are number states, $a^\dagger a
| n \rangle = n | n \rangle$ so that $\textrm{tr} \Delta_n = 1$.
In this case the classical bound in Eq. (\ref{trPQ}) becomes
\begin{equation}
p_n \leq \pi Q_{\mathrm{max}} = p_b ,
\end{equation}
which is actually independent of the outcome $n$.

A readily demonstration of the nonclassical nature of the
photon-number measurement is provided when $n=1$ and the
measured state is the one-photon state $|n =1 \rangle$. In
such a case $p_1 = 1$,
\begin{equation}
\label{Q1}
Q(\alpha) = \frac{|\alpha |^2}{\pi} \exp(-|\alpha |^2),
\quad \pi Q_{\mathrm{max}} = \frac{1}{e},
\end{equation}
where the maximum occurs for $|\alpha |=1$. Thus we have that
\begin{equation}
p_1 = 1 > \pi Q_{\mathrm{max}} \textrm{tr} \Delta_1  = \frac{1}{e},
\end{equation}
so that the measurement is nonclassical and the classical upper
bound is surpassed by 172 \%, since $(p_1 - p_b)/p_b = 1.72$.

\subsection{Quadrature measurements}

Concerning quadrature measurements (implemented in practice
by homodyne detection \cite{MSZ}) we have $\Delta_x = | x
\rangle \langle x |$, where $| x \rangle$ are the eigenstates
of the quadrature operator
\begin{equation}
\label{qo}
X = \frac{1}{2} \left ( a^\dagger + a \right ), \quad
X | x \rangle = x| x \rangle,
\end{equation}
being the optical analog of mechanical position or linear
momentum. In this case $\textrm{tr} \Delta_x$ is not finite
since $| x \rangle$ are not normalizable $\langle x | x^\prime
\rangle = \delta (x-x^\prime)$.

In order to avoid this difficulty we can appreciate that
the $P$ representative of $| x \rangle \langle x |$,
$P_x (\alpha = x^\prime + i y^\prime)$, does not depend on
$y^\prime$. This is an observable property, for example, via
the independence of statistics under displacements of the
measured state along this coordinate. Thus we can rearrange
Eq. (\ref{pm}) in the form
\begin{equation}
p_x = \pi \int dx^\prime P_x (x^\prime) \tilde{Q} (x^\prime)  ,
\quad
\tilde{Q} (x^\prime ) =  \int dy^\prime  Q (x^\prime ,y^\prime ),
\end{equation}
so that Eq. (\ref{trPQ}) is replaced by
\begin{equation}
p_x \leq \pi \tilde{Q}_\mathrm{max} \textrm{tr}_x \Delta_x ,
\end{equation}
where $\tilde{Q}_\mathrm{max}$ is the maximum of $\tilde{Q} (x)$
when $x$ is varied, and
\begin{equation}
\textrm{tr}_x \Delta_x = \int dx^\prime P_x (x^\prime) =
\int dx^\prime Q_x (x^\prime) ,
\end{equation}
where $P_x$, and $Q_x$ are the representatives of $\Delta_x$,
with
\begin{eqnarray}
\label{Qx}
Q_x (x^\prime ) &=& \frac{1}{\pi} |\langle x | \alpha = x^\prime
+ i y^\prime \rangle |^2 \nonumber\\
&=& \frac{1}{\pi} \sqrt{\frac{2}{\pi}}
\exp \left [ - 2 (x-x^\prime )^2 \right ] .
\end{eqnarray}
This leads to $\textrm{tr}_x \Delta_x = 1/\pi$ and to the
classical upper bound,
\begin{equation}
\label{bpx}
p_x \leq \tilde{Q}_\mathrm{max} = p_b ,
\end{equation}
that does not depend on the outcome $x$.

\subsubsection{Thermal-chaotic state}

In order to look for violations of bound (\ref{bpx}) let
us consider that the measured state $\rho$ is the thermal-chaotic
state whose expression in photon-number basis is
\begin{equation}
\label{tc}
\rho_{tc} = \left ( 1 - \xi \right ) \sum_{n=0}^\infty
\xi^n | n \rangle \langle n | ,
\end{equation}
where $\xi$ is a real parameter with $0 \leq \xi < 1$. These
states describe most classical light sources. The mean
number of photons $n_{tc}$ and the quadrature variance are
\begin{equation}
\label{ntc}
n_{tc} =\frac{\xi}{1-\xi},
\quad
\left ( \Delta X \right )^2 = \frac{1}{4} \left ( 1+ 2 n_{tc}
\right ) ,
\end{equation}
while the $Q$ and $\tilde{Q}$ functions are
\begin{eqnarray}
\label{Qtc}
& Q (\alpha ) = \frac{1}{\pi (n_{tc} +1)} \exp \left ( -
\frac{| \alpha |^2}{n_{tc} +1} \right ) , & \nonumber \\
& \tilde{Q} (x^\prime )= \frac{1}{\sqrt{\pi (n_{tc} +1)}}
\exp \left ( -  \frac{x^{\prime 2}}{n_{tc} +1} \right ) , &
\end{eqnarray}
so that the upper bound in Eq. (\ref{bpx}) reads as
\begin{equation}
p_b = \frac{1}{\sqrt{\pi (n_{tc} +1)}} .
\end{equation}
The statistics of the quadrature measurement $p_x = | \langle  x |
\rho_{tc} | x \rangle |^2$ is Gaussian
\begin{equation}
\label{eX}
p_x =\frac{1}{\sqrt{2 \pi} \Delta X} \exp \left [- \frac{x^2}{2
( \Delta X )^2} \right ] ,
\end{equation}
and the output most likely to break bound (\ref{trPQ}) is
$x=0$, since it maximizes $p_x$. This outcome will infringe the
bound provided that
\begin{equation}
p_ 0 = \frac{1}{\sqrt{\pi(n_{tc} + \frac{1}{2})}} >
p_b = \frac{1}{\sqrt{\pi(n_{tc} + 1)}},
\end{equation}
which holds for every $n_{tc}$. In particular for $n_{tc} =0$
(the vacuum state) we have $p_0 = 0.80$ and $p_b = 0.56$,
so that the classical upper bound is very clearly surpassed
by $100 (p_0 - p_b)/p_b = 43 \%$.

The outputs $x$ that contravene Eq. (\ref{bpx}) are all $x$
such that
\begin{equation}
\label{contra}
x^2 < \left ( \Delta X \right )^2 \ln \left [ 1 + \frac{1}{4
\left ( \Delta X \right )^2} \right ] .
\end{equation}
For $n_{tc}=0$ these are all $x$ in the interval $-0.42 \leq x \leq
0.42$, which occur with a 60 \% probability since
$\int_{-0.42}^{0.42} p_x dx \simeq 0.60$.

\subsubsection{Squeezed vacuum}

As a further example, when the measured state is the squeezed
vacuum the quadrature statistics has again the Gaussian form
(\ref{eX}), being the $Q$ function
\begin{equation}
Q (x,y) = \frac{1}{\pi} \frac{4 \Delta X}{1+4 (\Delta X )^2}
\exp \left [ - \frac{2 x^2 + 8 ( \Delta X )^2 y^2}{1+4 (
\Delta X )^2} \right ],
\end{equation}
so that
\begin{equation}
\tilde{Q} (x) = \sqrt{\frac{2}{\pi [ 1+4 ( \Delta X )^2]}}
\exp \left [ - \frac{2 x^2}{1+4 ( \Delta X )^2} \right ],
\end{equation}
and
\begin{equation}
\tilde{Q}_\mathrm{max}= \sqrt{\frac{2}{\pi
[ 1+4 (\Delta X )^2]}}.
\end{equation}
The output most likely to break bound (\ref{bpx}) is
$x=0$, and in such a case the classical bound is surpassed
for all $\Delta X$, since
\begin{equation}
p_0 = \frac{1}{\sqrt{2 \pi} \Delta X} >
p_b = \sqrt{\frac{2}{\pi [ 1+4 (\Delta X )^2]}}.
\end{equation}
The most favorable situation is when $\Delta X$ is as small
as possible. For example, for $\Delta X = 0.1$ we have
$p_0 = 4.0$ and $\tilde{Q}_\mathrm{max} = 0.8$, so that
there is a percentage of  violation of $100(p_0 -p_b)/p_b =
400 \%$ approximately. The outputs $x$ that contravene
Eq. (\ref{bpx}) are given by Eq. (\ref{contra}), which for
$\Delta X = 0.1$ is the interval $-0.18 \leq x \leq 0.18$,
that represents the 93 \% of all outcomes since
$\int_{-0.18}^{0.18} p_x dx \simeq 0.93$.

\section{Nonclassical states}

In this section we derive classical bounds disclosing
nonclassical measured states. They can be derived from the
last equality in  Eq. (\ref{pm}) by considering that for
classical states, i. e., for ordinary non-negative functions
$P (\alpha) \geq 0$, we get
\begin{equation}
P (\alpha) Q_m (\alpha) \leq P (\alpha) Q_{m,\mathrm{max}},
\end{equation}
where $Q_{m,\mathrm{max}}$ is the maximum of $Q_m (\alpha)$.
Applying this to the last equality in Eq. (\ref{pm}) and taking
into account Eq. (\ref{nor}), we get the following upper bound
for $p_m$,
\begin{equation}
\label{trQP}
p_m  \leq \pi Q_{m,\mathrm{max}} ,
\end{equation}
that holds for every $P (\alpha)$ compatible with classical
physics. If this condition is violated for any $m$ the state is
not classical. The bound becomes an equality when the measured
state is the coherent state $| \alpha_{m,\mathrm{max}} \rangle$
with $Q_{m,\mathrm{max}} = Q_m (\alpha_{m,\mathrm{max}})$.

Moreover, when the measured state is pure $\rho = | \psi \rangle
\langle \psi |$ we have that $| \psi \rangle$ is nonclassical if
and only if there is at least a measurement for which the classical
bound (\ref{trQP}) is violated. The violation of bound (\ref{trQP})
is clearly a sufficient condition. This is also necessary since
for every nonclassical $| \psi \rangle$ we can consider a POVM with
$\Delta_0 = | \psi \rangle \langle \psi |$. In such a case the upper
bound (\ref{trQP}) is surpassed because $p_0 = 1$ while $ \pi
Q_{0,\mathrm{max}} = | \langle \alpha_\mathrm{max} | \psi \rangle |^2
< 1$, since otherwise the equality $| \langle \alpha_\mathrm{max} |
\psi \rangle | = 1$ would imply that $|\psi \rangle$ is a coherent
state and thus classical.

Note that the POVM $\Delta_\alpha = | \alpha \rangle \langle \alpha |
 /\pi$ defined by the coherent states $| \alpha \rangle$ (implemented
in practice by double homodyne and heterodyne detection) are useless
for the detection of nonclassical states, since $\pi Q_{\alpha,
\mathrm{max}} = 1$ and bound (\ref{trQP}) becomes trivial
$p_\alpha \leq 1$.

For the sake of illustration we particularize this approach to
two meaningful practical situations. These are photon-number (Sec. III A) and quadrature measurements (Sec. III B).
Then we apply them to different measured states (Sec. III C).

\subsection{Photon-number measurement}

For photon-number measurements $\Delta_n = | n \rangle \langle n |$
the $Q$ function is
\begin{equation}
Q_n (\alpha)=\frac{1}{\pi}\exp(-|\alpha|^2)\frac{
|\alpha|^{2n}}{n!} ,
\end{equation}
and the maximum occurs at $|\alpha|=\sqrt{n}$:
\begin{equation}
Q_{n,\mathrm{max}} = \frac{1}{\pi}\exp(-n)\frac{n^{n}}{n!} .
\end{equation}
If the measured state is classical, the photon-number statistics
$p_n$ is thus bounded by
\begin{equation}
\label{cpr}
p_{n} \leq \exp(-n)\frac{n^{n}}{n!} = p_{b,n}.
\end{equation}
This bound was previously derived in Ref. \cite{MH}.
The upper bound $p_{b,n}$ is the probability of detecting $n$
photons in the coherent state $| \alpha \rangle$ with $|
\alpha | = \sqrt{n}$, which is the classical state for which
$p_n$ is maximum. In Fig. 1 we have represented $p_{b,n}$ as a
function of $n$ showing that for large $n$ it decays as
$p_{b,n} \simeq 1/\sqrt{2 \pi n}$ approximately, in agreement
with the Stirling approximation $n! \simeq \sqrt{2 \pi n} n^n
\exp (-n)$.

\begin{figure}
\begin{center}
\includegraphics[width=6cm]{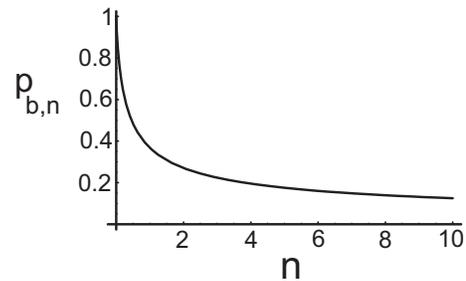}
\end{center}
\caption{Classical upper bound (\ref{cpr}) for the
probability of detecting $n$ photons on a classical state.}
\end{figure}

\subsubsection{Independence of sub-Poissonian statistics}

We can show that the nonclassical criterion on photon-number
measurements (\ref{cpr}) is independent of sub-Poissonian
statistics. The deviation from Poissonian statistics is
usually assessed by the Mandel parameter \cite{MSZ}
\begin{equation}
Q_M =\frac{(\Delta n)^2}{\langle n\rangle}-1 .
\end{equation}

The independence holds because:
(i) there are sub-Poissonian states that satisfy the classical
bounds (\ref{cpr}) for all $n$, and (ii) there are
super-Poissonian states that infringe them. To show this let
us consider the state in the number basis,
\begin{equation}
\rho = (1-p) | 0 \rangle \langle 0 | + p | N \rangle \langle N |,
\end{equation}
with $0 < p \leq 1$ so that
\begin{equation}
\langle n \rangle = pN , \quad (\Delta n)^2 = N^2 p(1-p) ,
\end{equation}
and
\begin{equation}
Q_M = N(1-p)-1 .
\end{equation}
(i) For $N=1$ this state is sub-Poissonian for all $p$ with $Q_M
= -p <0$, and satisfies the classical upper bounds (\ref{cpr})
for all $n$ when $p \leq 1/e$. (ii) For $p < (N-1)/N$ the state
is super-Poissonian since $Q_M > 0$, and infringes bound
(\ref{cpr}) when $p > p_{b,N}$. These two requirements are
compatible since it holds that $(N-1)/N > p_{b,N}$ for all $N>1$.

\subsection{Quadrature measurement}

As a further example we may consider the measurement of the
quadrature $X$ in Eq. (\ref{qo}), so that $\Delta_x = | x
\rangle \langle x|$. In such a case, from Eq. (\ref{Qx}) we get
$\pi Q_{x,\mathrm{max}} = \sqrt{2/\pi}$, and the classical upper
bound for the statistics $p_x$ of the quadrature measurement is
\begin{equation}
\label{ubpx}
p_x \leq  \sqrt{\frac{2}{\pi}} = p_b ,
\end{equation}
that does not depend on the output $x$. The maximum is obtained
for a coherent state $|\alpha \rangle$ with $(\alpha + \alpha^\ast
)/2 = x$. For states with Gaussian $p_x$ the infringement of
Eq. (\ref{ubpx}) is equivalent to squeezing of quadrature $X$
since the maximum of $p_x$ is $1/(\Delta X \sqrt{2 \pi})$ and
\begin{equation}
\frac{1}{\Delta X \sqrt{2 \pi}} > \sqrt{\frac{2}{\pi}}
\longrightarrow
\Delta X < \frac{1}{2} = \Delta X_{\mathrm{vacuum}} .
\end{equation}
For non Gaussian $p_x$ the situation can be different as
shown below.

This example is interesting since quadrature measurements
are more experimentally feasible than number measurements. For a
further discussion about the quantum-classical relation in
terms of quadrature distributions see Ref. \cite{MMT}.

\subsection{Examples}

Let us consider some meaningful simple examples of states
violating the classical upper bounds (\ref{cpr}) and (\ref{ubpx}).

\subsubsection{Incoherent superposition of thermal and number}

As a feasible state that can infringe Eq. (\ref{cpr}) let us
consider the incoherent superposition of the thermal-chaotic
state in Eq. (\ref{tc}) and the photon number state $| n_0
\rangle$
\begin{equation}
\label{rtcn}
\rho = p \rho_{tc} + (1-p) | n_0  \rangle \langle n_0  |,
\end{equation}
where $0  \leq p \leq 1$, leading to a photon-number statistics
\begin{equation}
\label{ipn}
p_n = p (1-\xi) \xi^n + (1-p) \delta_{n,n_0}.
\end{equation}
For example, for $p=0.5$, $n_0 =1$, and $n_{tc}=9$ the
probability of detecting a single photon is $p_{1}=0.545$,
while the upper bound in Eq. (\ref{cpr}) for $n=1$ is $p_{b,1}
= 1/e = 0.368$, so we have a clear infringement of the
classical condition (\ref{cpr}) by $100 (p_1 -p_{b,1} )/p_{b,1}
= 48 \%$.

In this case the photon-number distribution (\ref{ipn}) is
highly super-Poissonian with $Q_M = 11.2$. Furthermore, we can
easily show that there is no quadrature squeezing since for
quadrature operators
\begin{equation}
\label{rqo}
X_\theta = \frac{1}{2} \left [ a^\dagger \exp (-i\theta)
+ a \exp (i\theta) \right ],
\end{equation}
we have in state (\ref{rtcn}) that $\langle X_\theta \rangle
=0$ and
\begin{equation}
\left ( \Delta X_\theta \right )^2 = \frac{1}{2} \langle n \rangle
+ \frac{1}{4} = \frac{1}{2} \left [ p n_{tc} + (1-p) n_0 \right ] +
\frac{1}{4}  .
\end{equation}
For the above parameters, $n_0 = 1$, $p=0.5$, and $n_{tc} = 9$, we
get $( \Delta X_\theta )^2 = 11/4$ for all $\theta$, which is
far above the upper limit for squeezing $( \Delta
X_\theta )^2_{\mathrm{vacuum}} = 1/4$. Finally, it can be
appreciated that there are no oscillations in the photon-number
distribution.

For state (\ref{rtcn}) the origin of nonclassical behavior
is that $P( \alpha)$ is always more singular than a delta
function for all $p \neq 1$. This is because
\begin{equation}
\label{sup}
P (\alpha) = p P_{tc} (\alpha) + (1-p) P_{n_0} (\alpha) ,
\end{equation}
where
\begin{equation}
P_{tc} (\alpha) = \frac{1}{\pi n_{tc}} \exp \left ( -
\frac{|\alpha |^2 }{n_{tc}} \right ) ,
\end{equation}
and
\begin{equation}
P_{n_0 = 1} (\alpha) = \left ( 1 + \frac{\partial}{\partial
\alpha} \frac{\partial}{\partial \alpha^\ast } \right )
\delta^{(2)} (\alpha ) .
\end{equation}
Thus $P(\alpha)$ is more singular than a delta function since
otherwise we would be able to express $P_{n_0 =1}(\alpha)$
as a linear combination of two ordinary functions.

\subsubsection{Photon-added thermal state}

The previous states are associated with $P$ representatives more
singular than a delta function. Next we consider states with
nonsingular $P(\alpha)$ function taking negative values. This is the case of
the single-photon-added thermal states that, in the photon-number
basis, read as \cite{AT}
\begin{equation}
\label{pats}
\rho_1 = (1 - \xi) a^\dagger \rho_{tc} a = (1 - \xi)^2
\sum_{n=1}^\infty \xi^{n-1} n | n \rangle
\langle n | ,
\end{equation}
where $\rho_{tc}$, $\xi$ and $n_{tc}$ are in Eqs. (\ref{tc}) and
(\ref{ntc}) respectively.

The $P$ representative is well behaved but nonpositive,
\begin{equation}
P(\alpha) = \frac{1}{\pi n^3_{tc}} \left [ \left ( n_{tc} + 1
\right ) | \alpha |^2 - n_{tc}  \right ] \exp \left ( - |
\alpha |^2 / n_{tc} \right ),
\end{equation}
and the photon-number statistics is
\begin{equation}
\label{ips}
p_n = (1 - \xi)^2 \xi^{n-1} n.
\end{equation}

For example, for $n=1,2$ the classical upper bounds (\ref{cpr})
are surpassed provided that
\begin{eqnarray}
& p_1 > p_{b,1} = 1/e \leftrightarrow n_{tc} < \sqrt{e}-1 =
0.65 , & \nonumber \\
& p_2 > p_{b,2} = 2/e^2 \leftrightarrow
0.30 \leq n_{tc} \leq 0.82 . &
\end{eqnarray}
Let us show that this nonclassical behavior is independent
of other nonclassical features. There is no quadrature
squeezing since for the rotated quadrature operators $X_\theta$
in Eq. (\ref{rqo}) we have $(\Delta X_\theta )^2 = (3 + 4 n_{tc})/4
> 1/4$ for all $\theta$. Also, it can be appreciated in
Eq. (\ref{ips}) that there is no photon-number oscillations.
Finally, the Mandel parameter is
\begin{equation}
Q_M = \frac{2 n_{tc}^2 - 1}{2 n_{tc} + 1} ,
\end{equation}
so we get super-Poissonian statistics for all $n_{tc} >
1/\sqrt{2} = 0.71$.

Therefore, the states with $0.71 \leq n_{tc} \leq 0.82$ are
nonclassical since $p_2 > p_{b,2}$, although they have
super-Poissonian statistics, present no squeezing, and have
no oscillatory statistics.

\subsubsection{Coherent superposition of coherent states}

Another interesting example is provided by the coherent
superposition of two coherent states with opposed complex
amplitude \cite{cqs} (referred to as even and odd
superpositions \cite{Do})
\begin{equation}
\label{scs}
|\alpha_{\pm} \rangle = N_{\pm} ( |\alpha\rangle \pm |
- \alpha \rangle ) ,
\end{equation}
with
\begin{equation}
N_+ = \frac{\exp(|\alpha|^2/2)}{2 \sqrt{\cosh (
|\alpha|^2 )}},
\quad
N_- = \frac{\exp(|\alpha|^2/2)}{2 \sqrt{\sinh (
|\alpha|^2 )}}.
\end{equation}
In this case the $P (\alpha )$ is a distribution
involving an infinite number of derivatives of the delta
function, since the normally-ordered characteristic function
is a real exponential. For definiteness let us focus
just on the even states $| \alpha_+ \rangle$.

For the even case $| \alpha_+ \rangle$ we have the following
photon-number statistics
\begin{equation}
\label{pmscs}
p_n = \left \{ \begin{array}{cl} \label{egato}
\frac{|\alpha|^{2n}}{n! \cosh (|\alpha|^2)}& \text{ for } n
\text{ even}, \\ 0& \text{ for } n \text{ odd} . \end{array}
\right .
\end{equation}
Numerically it can be easily seen that this is inconsistent
with the upper bound in Eq. (\ref{cpr}) for $|\alpha | \geq 0.64$.
Moreover, for $|\alpha| >>1$ we have $\cosh (|\alpha|^2) \simeq
\exp (|\alpha|^2)/2$ and (always for even $n$)
\begin{equation}
p_n \simeq 2 \frac{|\alpha|^{2n}}{n!} \exp (- |\alpha|^2) ,
\end{equation}
so that for $| \alpha |^2 = n $ we get that $p_n$ is twice the upper
bound $p_{b,n}$ in Eq. (\ref{cpr}).

Concerning quadrature measurements, let us consider states with
purely imaginary complex amplitude $\alpha = \pm i | \alpha |$
that have the following quadrature statistics:
\begin{equation}
\label{pxsc}
p_x = 4 N_+^2 \sqrt{\frac{2}{\pi}} \cos^2 ( 2 | \alpha | x )
\exp \left ( - 2 x^2 \right ) .
\end{equation}
For every $|\alpha|$ the maximum of $p_x$ holds for $x=0$
being $p_0 = 4 N_+^2 \sqrt{2/\pi}$. In Fig. 2 (dashed line) we
have represented the relative amount of violation of Eq. (\ref{ubpx}),
$100(p_0 - p_{b})/p_{b}$, as a function of $| \alpha |$, showing
a 100 \% violation for large $| \alpha |$. This is because for
$|\alpha |>>1$ we have $N_+^2 \simeq 1/2$ so that $p_0 \simeq  2
\sqrt{2 /\pi}$, which is twice the classical upper bound $p_b =
\sqrt{2 /\pi}$. In Fig. 2 this is also compared with the percentage
of squeezing in the same state (solid line).

\begin{figure}
\begin{center}
\includegraphics[width=6cm]{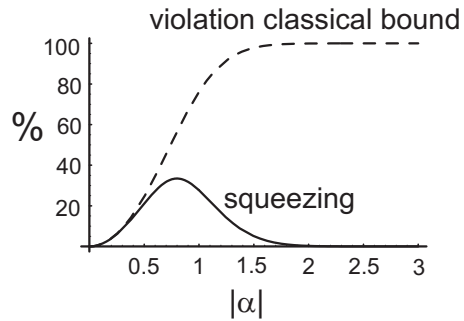}
\end{center}
\caption{Percentage $100( p_0 - p_{b})/p_{b}$ of maximum
violation of the classical bound (\ref{ubpx}) (dashed line) and
percentage of squeezing $100(1- 2 \Delta X_{\theta, \mathrm{min}})$
in Eq. (\ref{pesq}) (solid line) for the even state (\ref{scs})
as functions of $| \alpha |$.}
\end{figure}

Next we show that the even states infringe classical bounds with
super-Poissonian photon-number statistics and with negligible
quadrature squeezing. Concerning photon-number statistics we have
\begin{eqnarray}
& \langle n \rangle = | \alpha |^2 \tanh ( | \alpha |^2 ), &
\nonumber \\
& \langle n^2 \rangle = | \alpha |^4  + | \alpha |^2
\tanh ( | \alpha |^2 ),  &
\end{eqnarray}
so that
\begin{equation}
Q_M = \frac{2 | \alpha |^2}{\sinh \left ( 2 | \alpha |^2 \right )} ,
\end{equation}
and these states are always super-Poissonian (unless $\alpha = 0)$.

Concerning quadrature squeezing, the minimum uncertainty for
rotated quadratures (\ref{rqo}) in the state $| \alpha_+ \rangle$
when $\theta$ is varied is
\begin{equation}
(\Delta X_\theta )^2_{\mathrm{min}} = \frac{1}{4} \left (
2 |\alpha|^2\tanh|\alpha|^2 - 2 |\alpha|^2+1 \right ).
\end{equation}
The percentage of squeezing defined as
\begin{equation}
\label{pesq}
100 \frac{\Delta X_{\mathrm{vacuum}} - \Delta X_{\theta, \mathrm{min}}}
{\Delta X_{\mathrm{vacuum}}} = 100 \left ( 1- 2 \Delta X_{\theta,
\mathrm{min}} \right ),
\end{equation}
is represented in Fig. 2 as a function of $| \alpha |$. These states
present squeezing only for small $|\alpha |$ being negligible for
$|\alpha | > 2$. For instance, for $|\alpha | =3$ we have
$(\Delta X_{\theta, \mathrm{min}})^2 = 0.24999986$, which means a
fully negligible $2.75  \times 10^{-5} \%$ squeezing, while
Eqs. (\ref{cpr}) and (\ref{ubpx}) are infringed by a 100\% for the
same state.

At difference with the preceding examples in this case the 100 \%
violation of classical bounds for large $| \alpha |$ has a simple
explanation in terms of the oscillatory character of the statistics
(\ref{pmscs}) and (\ref{pxsc}). For large $| \alpha |$ the number
and quadrature statistics are the same of coherent states (that
would saturate the classical bounds) but maximally modulated.
Because of normalization, the vanishing terms must be compensated by
nonvanishing terms reaching twice the coherent-state values. This
factor of 2 leads to the 100\% violation of the classical bounds.

Let us note that the criteria presented in this work
reveal the nonclassical nature of these states for all $\alpha$,
but specially clearly for large $| \alpha|$. This is sharp
contrast with sub-Poissonian number statistics and quadrature
squeezing, that hold only for small $|\alpha|$, as illustrated
in Fig. 2 for example.

\section{Effect of imperfections}

One of the key features of this approach is that the data analysis
is reduced to minimum. This favours obtaining reliable results
from non ideal measurements affected by imperfections, such as
damping, finite efficiencies, or finite sampling. We stress that
this approach applies to any measurement, both ideal and
imperfect, so that experimental imperfections can be always
embodied into the measuring POVM. Nevertheless, since imperfections
usually deteriorate nonclassical properties it is reasonable to
investigate their effect on the above nonclassical criteria.

\subsection{Inefficient detection}

For definiteness we consider real detectors affected by field
damping (with bath at zero temperature) and finite quantum
efficiency, which  can be modeled by placing a beam splitter of
amplitude-transmission coefficient $t = \sqrt{\eta}$ in front
of a perfect detector, where $\eta \leq 1$ represents both losses
and efficiencies \cite{qeta,STG}.

\subsubsection{Nonclassical states}

The effect of the beam splitter for the detection of nonclassical
states can be easily accounted for by computing the measured state
after the beam splitter $\rho_t$ as
\begin{equation}
\label{ims}
\rho_t = \int d^2 \alpha P(\alpha ) | t \alpha \rangle \langle t
\alpha | ,
\end{equation}
where $P(\alpha )$ is the $P$ function of the measured state. The
measured statistics becomes
\begin{equation}
p_{t,m} = \textrm{tr}\left ( \Delta_m \rho_t \right ) = \pi \int d^2
\alpha P (\alpha) Q_m (t \alpha) .
\end{equation}
Since the maximum of $Q_m (t \alpha)$ when $\alpha$ is varied is
the same as the maximum of $Q_m (\alpha)$, there is no change in
the classical upper bound in the right-hand side of Eq. (\ref{trQP}).

Nevertheless, imperfections affect the statistics replacing $p_m$
by $p_{t,m}$ in the left-hand side of Eq. (\ref{trQP}). This can be
easily seen, for example, for photon-added thermal states (\ref{pats}).
For inefficient detection the photon-number statistics of ideal
case (\ref{ips}) for $n=1$ is replaced by
\begin{equation}
\label{pt12}
p_{t,1} = \eta \frac{1 + 2 n_{tc} - \eta n_{tc}}{ \left ( 1+
\eta n_{tc} \right )^3} .
\end{equation}
In Fig. 3 we have represented $p_{t,1}$ as a function of $\eta$ for
$n_{tc} = 0.7$. Decreasing $\eta$ from $\eta =1$ increases $p_{t,1}$,
leading to break the classical bound (\ref{cpr}) in the interval
$0.30 \leq \eta \leq 0.89$.

It is worth pointing out that, at difference with other nonclassical
tests, where imperfections degrade nonclassical behavior \cite{STG},
in this case larger losses and decreasing efficiencies may favour
the observation of nonclassical behavior of measured states. This
noticeable effect arises because imperfections rearrange the
probability distribution $p_n$, so that with increasing imperfection
some probabilities may increase beyond the classical bounds, as is
the case of $p_{t,1}$ in this example.

\begin{figure}
\begin{center}
\includegraphics[width=6cm]{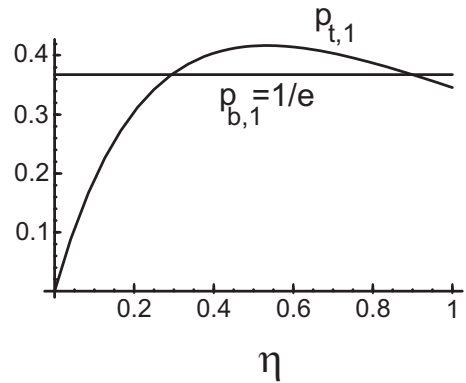}
\end{center}
\caption{$p_{t,1}$ in Eq. (\ref{pt12}) as a function of $\eta$
for $n_{tc} = 0.7$.}
\end{figure}

\subsubsection{Nonclassical measurements}

For nonclassical measurements we can follow two different
strategies: (i) we can address the nonclassical behavior of the ideal
POVM $\Delta_m$ associated to $\eta = 1$. This is the analog of the
preceding subsection where we investigated the nonclassical
properties of the input state before being affected by imperfections.
(ii) Alternatively, as mentioned above we can examine the nonclassical
behavior of the effective POVM $\tilde{\Delta}_m$ embodying all
imperfections as part of the measuring scheme.

(i) Concerning the nonclassical behavior of the ideal
POVM $\Delta_m$ we can compute the effect of inefficiencies as
\begin{equation}
p_{t,m} = \frac{1}{\pi} \int d^2 \alpha d^2 \beta \;
\mbox{}_s \langle \alpha | \mbox{}_a \langle \beta | T \rho \otimes
\rho_0 T^\dagger | \beta \rangle_a | \alpha \rangle_s  P_m (\alpha) ,
\end{equation}
where $| \alpha \rangle_s$ are coherent states in the signal
mode (the mode of the measured state $\rho$) and $| \beta
\rangle_a$ are coherent states in the auxiliary mode (the other
input port of the beam splitter assumed in the vacuum state
$\rho_0$), $P_m (\alpha)$ is the $P$ representative
of the ideal measurement $\Delta_m$, and $T$ is the unitary
transformation describing the effect of the beam splitter, with
\begin{equation}
T^\dagger | \beta \rangle_a | \alpha \rangle_s = | t \beta + r
\alpha \rangle_a | t \alpha - r \beta \rangle_s ,
\end{equation}
being $r = \sqrt{1-t^2}$. This leads to the following form for
the statistics
\begin{equation}
\label{ie}
p_{t,m} = \pi \int d^2 \alpha \tilde{Q} (\alpha) P_m (\alpha),
\end{equation}
where $\tilde{Q} (\alpha)$ is defined here as
\begin{equation}
\tilde{Q} (\alpha) = \int d^2 \beta Q_0 (t \beta + r \alpha)
Q (t \alpha - r \beta) ,
\end{equation}
with $Q_0$ and $Q$ being the $Q$ representatives of $\rho_0$ and $\rho$,
respectively. Note that in the ideal case $\eta=1$ ($t=1$, $r=0$)
$\tilde{Q} (\alpha)$ is the $Q$ function of the measured state
$\tilde{Q} (\alpha) = Q(\alpha)$. From Eq. (\ref{ie}) we can
derive the classical upper bound
\begin{equation}
\label{clupbo}
p_{t,m} \leq \pi \tilde{Q}_{\mathrm{max}} \textrm{tr} \Delta_m ,
\end{equation}
which holds for classical measurements with $P_m (\alpha) \geq 0$.
We can appreciate that  finite quantum efficiencies modify the
classical upper bounds in comparison with the ideal detection
in Eq. (\ref{trPQ}) by replacing $Q_{\mathrm{max}}$ by
$\tilde{Q}_{\mathrm{max}}$.

Let us illustrate this analysis with the example
where the ideal POVM is one-photon detection,
$\Delta_1 = | 1 \rangle \langle 1 |$ in the number basis,
and the measured state is the one-photon state $| 1 \rangle$.
The case $\eta =1$ was considered in Sec. II A above.
When $\eta \leq 1$ we have
\begin{equation}
p_{t,1} = \eta, \quad \textrm{tr} \Delta_1 = 1 ,
\end{equation}
and
\begin{equation}
\tilde{Q} (\alpha) = \frac{1}{\pi} \left [ \eta \left (
| \alpha |^2 - 1 \right ) + 1 \right ] \exp \left ( - |
\alpha |^2 \right ) ,
\end{equation}
so that
\begin{equation}
\tilde{Q}_{\mathrm{max}} = \frac{\eta}{\pi} \exp \left (
- \frac{2 \eta -1}{\eta} \right ) ,
\end{equation}
and the violation of the classical upper bound (\ref{clupbo})
occurs provided that
\begin{equation}
\exp \left (  - \frac{2 \eta -1}{\eta} \right ) < 1
\longleftrightarrow \eta > \frac{1}{2} .
\end{equation}
Therefore, in this example the nonclassical behavior of the
ideal measurement is disclosed provided that the quantum
efficiencies are above 50 \%.

(ii) Alternatively, if we embody decaying mechanisms
and inefficiencies in the effective POVM $\tilde{\Delta}_m$ we
get from Eq. (\ref{trPQ})
\begin{equation}
\label{ibg}
p_{t,m} \leq \pi Q_\mathrm{max} \textrm{tr} \tilde{\Delta}_m,
\end{equation}
where $Q_\mathrm{max}$ is the maximum of the $Q$ function
of the measured state $\rho$. We can compute $ \textrm{tr}
\tilde{\Delta}_m$ taking into account the effect of the beam
splitter as in Eq. (\ref{ims}):
\begin{eqnarray}
 \textrm{tr} \tilde{\Delta}_m &=&\frac{1}{\pi} \int d^2
\alpha \langle \alpha | \tilde{\Delta}_m  | \alpha \rangle \nonumber\\
&=&\frac{1}{\pi} \int d^2 \alpha \langle t \alpha | \Delta_m  |
t \alpha \rangle = \frac{1}{t^2}  \textrm{tr} \Delta_m  ,
\end{eqnarray}
leading to
\begin{equation}
\label{ib}
p_{t,m} \leq \pi Q_\mathrm{max} \frac{1}{\eta} \textrm{tr}
\Delta_m .
\end{equation}
We can appreciate that the effect of imperfections is
simply expressed by increasing the classical upper bound by
a factor of $1/\eta$.

Let us illustrate this approach with the same example of
inefficient one-photon detection, (ideal POVM $\Delta_1 = | 1
\rangle \langle 1 |$ and a one-photon state $| 1 \rangle$) so
that after Eq. (\ref{Q1}) the classical bound (\ref{ib}) is
\begin{equation}
p_{t,1} = \eta \leq \frac{1}{\eta e} .
\end{equation}
Thus the effective POVM $\tilde{\Delta}_m$ shows
nonclassical behavior when $\eta > 1/\sqrt{e} = 0.6065$.

We can appreciate that the two approaches (i) and
(ii) lead to two different bounds, (\ref{clupbo}) and
(\ref{ib}), as clearly illustrated by the example of
one-photon detection. We stress that this difference is
natural since the classical bound (\ref{clupbo}) is
sensitive to the nonclassicality of the $P$ representative
of the ideal POVM $\Delta_m$, while bound (\ref{ib})
is sensitive to the nonclassical character of the $P$
representative of the effective POVM $\tilde{\Delta}_m$.

\subsection{Finite sampling}

When the number of measurements $N$ is finite, the probability
$p_m$ becomes an statistical variable that can be expressed as
$p_m (N) = k/N$, where the integer $k$ is the number of outcomes
$m$ after $N$ trials. (This analysis applies both to detection
of nonclassical states and measurements.) The dichotomic character
of the measurement (outcome $m$ with probability $p_m$ and
outcome not $m$ with probability $1-p_m$) implies that $k$ follows
the binomial distribution
\begin{equation}
\mathcal{P}_k (N) = \pmatrix{N \cr k} p_m^k (1-p_m)^{N-k} ,
\end{equation}
so we have
\begin{eqnarray}
\langle p_m (N) \rangle &=& \frac{\langle k \rangle}{N} = p_m,\nonumber \\
\Delta p_m (N) &=& \frac{\Delta k}{N} = \sqrt{\frac{p_m (1-p_m)}{N}}.
\end{eqnarray}
For all the above examples we have roughly $p_m \simeq 0.5$,
so that for $N \simeq 100$ we have $\Delta p_m /p_m \simeq
0.1$. Thus, even for moderate number of trials, the uncertainty
caused by finite sampling is clearly below the amount
of violation of classical upper bounds $p_{b,m}$, since
$(p_m - p_{b,m})/p_{b,m}$ is at least five time larger than
$\Delta p_m /p_m$ in the above examples.

\section{Two-mode observables}

The above single-mode approach in Eq. (\ref{pm}) can be easily
generalized to two-mode observables by expressing the statistics
$p_m = \mathrm{tr}( \rho \Delta_m)$ as
\begin{equation}
p_m = \pi^2 \int d^2 \alpha d^2 \beta P(\alpha,\beta)
Q_m (\alpha,\beta) ,
\end{equation}
and
\begin{equation}
p_m = \pi^2 \int d^2 \alpha d^2 \beta
P_m (\alpha,\beta) Q (\alpha,\beta) ,
\end{equation}
where $P$, and $Q$ are the two-mode phase-space representatives for
the measured state $\rho$
\begin{eqnarray}
\rho = \int d^2 \alpha d^2 \beta P(\alpha,\beta)| \alpha , \beta
\rangle \langle \alpha , \beta | \\
Q(\alpha,\beta) = \frac{1}{\pi^2} \langle \alpha , \beta | \rho
| \alpha , \beta \rangle ,
\end{eqnarray}
$P_m$, and $Q_m$ refer to the corresponding representatives of the
POVM $\Delta_m$, and $| \alpha, \beta \rangle$ are two-mode coherent
states. The parameter $m$ represents all the indices necessary to
label the outcomes. The maxima of $Q (\alpha,\beta)$ and $Q_m
(\alpha,\beta)$ provide suitable upper bounds for the statistics
of classical measurements and states, respectively.

In this regard we note that for bipartite systems nonclassical
$P (\alpha, \beta)$ is a necessary condition for entanglement
\cite{SV}. For definiteness we focus on the nonclassical behavior
of states. The analysis of nonclassical measurements would be
analogous.

\subsection{Nonclassical states by photon-number detection}

For the case of joint two-mode photon-number detection we get that
for classical states the joint probability of detecting $n_1$ and
$n_2$ photons is simply bounded by the product of the one-mode upper
bounds
\begin{equation}
p_{n_1 , n_2} \leq p_{b,n_1} p_{b,n_2} = \exp [ - (n_1 + n_2 ) ]
\frac{n_1^{n_1} n_2^{n_2}}{n_1! n_2!} .
\end{equation}
The maximum for fixed $n_1 + n_2$ occurs when $n_1 =0$ or $n_2
= 0$ while the minimum occurs for coincident outputs $n_1 = n_2$.

For the total number $n = n_1 + n_2$ the statistics is given by
\begin{equation}
p_{n} = \pi^2 \int d^2 \alpha d^2 \beta P(\alpha,\beta)
\sum_{m=0}^n Q_{m,n-m} (\alpha,\beta) ,
\end{equation}
where
\begin{equation}
Q_{n_1,n_2} (\alpha,\beta) = \frac{1}{\pi^2} \left | \langle
\alpha | n_1 \rangle \right |^2 \left | \langle \beta | n_2
\rangle \right |^2 ,
\end{equation}
$| n_{1,2} \rangle$ being number states in the corresponding
modes. It can be easily seen that
\begin{equation}
\sum_{m=0}^n Q_{m,n-m} (\alpha,\beta) = \frac{1}{\pi^2}
\frac{|\gamma |^{2n}}{n!} \exp (- | \gamma |^2 ) ,
\end{equation}
with $| \gamma |^2 = | \alpha |^2 + | \beta |^2$. For fixed $n$
the maximum occurs for $| \gamma |^2 = n$ so that
\begin{equation}
\sum_{m=0}^n Q_{m,n-m} (\alpha,\beta) \leq \frac{1}{\pi^2}
\frac{n^{n}}{n!} \exp (- n ) ,
\end{equation}
and
\begin{equation}
p_n \leq \frac{n^{n}}{n!} \exp (- n ) ,
\end{equation}
which is equal to the single-mode counterpart (\ref{cpr}).

\subsection{Nonclassical states by quadrature-difference measurement}

Let us consider the measurement of the quadrature difference
$X = X_1 - X_2$, where $X_{1,2}$ represent the same quadrature
operator in each mode, which is described by the POVM
\begin{equation}
\Delta_x = \int d x^\prime | x + x^\prime \rangle_1
\langle x + x^\prime | \otimes | x^\prime \rangle_2
\langle x^\prime | ,
\end{equation}
where $ | x \rangle_j$ are the eigenstates of $X_j$, with $j=1,2$.
In this case we have
\begin{equation}
Q_x (\alpha_1, \alpha_2) = \frac{1}{\pi^2 \sqrt{\pi}}
\exp \left [ - (x - x_1 + x_2 )^2 \right ],
\end{equation}
where $x_j$ is the real part of $\alpha_j$, so that
\begin{equation}
\pi^2 Q_{x, \mathrm{max}} = \frac{1}{\sqrt{\pi}}
\end{equation}
and the classical bound is
\begin{equation}
\label{ocb}
p_x \leq  \frac{1}{\sqrt{\pi}} = p_b .
\end{equation}
Note that the two-mode bound does not depend on the outcome $x$,
being lower than the single-mode counterpart (\ref{ubpx}).
For states with Gaussian $p_x$ the violation of this bound is
equivalent to $\Delta X < 1/\sqrt{2}$. This is equivalent to
two-mode squeezing since for pairs of coherent states it
holds $\Delta X = 1/\sqrt{2}$.

\subsection{Example: Two-mode squeezed vacuum}

To illustrate these two-mode classical bounds let us consider
that the measured state is a two-mode squeezed vacuum, that in
the photon-number basis reads,
\begin{equation}
| \zeta \rangle = \sqrt{1-\zeta^2} \sum_{n=0}^\infty \zeta^n
| n \rangle_1 | n \rangle_2 ,
\end{equation}
where we have assumed real parameter $\zeta$ without loss of
generality. The statistic of the joint number $p_{n,n}$
and the total number $p_{2n}$ are
\begin{equation}
p_{n,n} = p_{2n}=\left ( 1 - \zeta^2 \right ) \zeta^{2n} ,
\end{equation}
while the statistics of the quadrature-difference $X = X_1 - X_2$
is Gaussian with
\begin{equation}
\left ( \Delta X \right )^2 =
\frac{1-\zeta}{2 (1+\zeta)} .
\end{equation}

Numerically we have found that the classical bound on
joint-number measurements is always violated $p_{n,n}>p^2_{b,n}$
for some $n$ when $\zeta >0.41$. The classical bound for total
number is never violated since $p_{2n} = p_{n,n} \leq p_{b,2n}$
for all $\zeta$.

More specifically, for $n=1$ we have  $p_{1,1} = ( 1 - \zeta^2 )
\zeta^2 > p^2_{b,1} = \exp (-2)=0.135$ for all $\zeta$ in the
interval $0.41 \leq \zeta \leq 0.91$. The maximum violation
occurs for $\zeta^2 = 1/2$ so that $p_{1,1}=0.25$, and there
is an 85 \% violation of the classical bound. For the total
number we have that the classical bound is not surpassed since
$p_{b,2}=2 \exp(-2) = 0.27$.

On the other hand the classical bound on quadrature difference is
always surpassed since the statistics is Gaussian and $\Delta X <
1/\sqrt{2}$ for all $\zeta$. For example, for $\zeta^2 = 1/2$ (this
is mean total number of photons $\langle n \rangle =2$) we have
$p_0 = 1.36$, while the classical upper bound in Eq. (\ref{ocb})
is $p_b = 0.56$, so that there is a percentage of violation $100 (p_0
-p_b)/p_b = 141 \%$ approximately. The outputs $x$ that contravene
Eq. (\ref{ocb}) are all $x$ in the interval $-0.39 \leq x \leq 0.39$,
which represent an 82\% probability since $\int_{-0.39}^{0.39}
p_x dx = 0.82$.

\section{Spin systems}

The above methods can be adapted to situations described by
finite-dimensional Hilbert spaces, exemplified by spin-$j$ systems.
This can be readily done in terms of SU(2) $Q$ and $P$ functions,
which are defined after the SU(2) coherent states $|j, \Omega
\rangle $ as \cite{ACGT}
\begin{equation}
\rho = \int d^2 \Omega P(\Omega) | j, \Omega \rangle
\langle j, \Omega | , \quad Q(\Omega) = \frac{2j+1}{4 \pi}
\langle j , \Omega | \rho | j, \Omega \rangle,
\end{equation}
with $d^2 \Omega = \sin \theta d \theta d \phi$, and
\begin{eqnarray}
\label{cs}
|j, \Omega \rangle = \sum_{m = -j}^j &\left ( \begin{array}{c} 2j
\cr m+j \end{array} \right )^{1/2}\sin^{j-m}
\left ( \frac{\theta}{2} \right ) \cos^{j+m} \left (
\frac{\theta}{2} \right ) \nonumber \\
&\cdot\exp [-i (j+m) \phi] |j,m \rangle ,
\end{eqnarray}
where $|j,m \rangle$ are the eigenstates of the spin component
$j_3$ with eigenvalue $m$, while $\pi \geq \theta \geq 0$ and
$\pi \geq \phi \geq - \pi$. The analog of Eq. (\ref{pm}) is
\begin{eqnarray}
p_m &=& \frac{4 \pi}{2j+1} \int d^2 \Omega P_m (\Omega)
Q (\Omega) \nonumber\\
&=& \frac{4 \pi}{2j+1} \int d^2 \Omega P (\Omega )
Q_m (\Omega)  ,
\end{eqnarray}
leading to the following bounds for classical measurements
\begin{equation}
\label{cbsu2a}
p_m \leq \frac{4 \pi}{2j+1} Q_{\mathrm{max}} \textrm{tr}
\Delta_m  ,
\end{equation}
(for finite-dimensional systems $\textrm{tr} \Delta_m$ is
always finite), while the upper bound for classical states
is
\begin{equation}
\label{cbsu2b}
p_m \leq \frac{4 \pi}{2j+1} Q_{m,\mathrm{max}} = p_{b,m} .
\end{equation}

By construction the SU(2) coherent states are classical both
as measured and measuring states. For the POVM
\begin{equation}
\label{su2cst1}
\Delta_\Omega = \frac{2j+1}{4 \pi} | j , \Omega \rangle
\langle j, \Omega |,
\quad
\textrm{tr} \Delta_\Omega = \frac{2j+1}{4 \pi} ,
\end{equation}
the upper bound (\ref{cbsu2a}) is satisfied for all measured
states since
\begin{equation}
p_\Omega = Q(\Omega) \leq  \frac{4 \pi}{2j+1} Q_{\mathrm{max}}
\textrm{tr} \Delta_\Omega  =  Q_{\mathrm{max}} ,
\end{equation}
where $Q (\Omega)$ is the SU(2) $Q$ function of the measured
state. Likewise, when the measured state is coherent $\rho =
| j , \Omega \rangle \langle j, \Omega |$ we have
\begin{equation}
\label{su2cst2}
p_m = \frac{4 \pi}{2j+1} Q_m (\Omega) ,
\end{equation}
where $Q_m (\Omega)$ is the SU(2) $Q$ function of $\Delta_m$, so
that (\ref{cbsu2b}) is satisfied by all measurements.

For the sake of illustration next we consider some
examples for the simplest cases $j=1/2$, and $j=1$.

\subsection{$j=1/2$}

For $j=1/2$ the SU(2) coherent states read, in the $~|j,m \rangle$
basis, as
\begin{eqnarray}
|1/2, \Omega \rangle &=& \sin \left ( \frac{\theta}{2} \right )
|1/2,-1/2 \rangle \nonumber\\
&+& \cos \left ( \frac{\theta}{2} \right )
\exp (-i \phi) |1/2, 1/2 \rangle .
\end{eqnarray}
Every state and POVM are of the form
\begin{equation}
\rho = \frac{1}{2} \left ( \sigma_0 + \bm{r} \cdot \bm{\sigma}
\right ), \quad
Q (\Omega ) = \frac{1}{4 \pi} \left ( 1 + \bm{r} \cdot
\bm{\Omega} \right ) ,
\end{equation}
\begin{equation}
\Delta_m = \lambda_m \left ( \sigma_0 + \bm{r}_m \cdot
\bm{\sigma} \right ), \quad
Q_m (\Omega ) = \frac{\lambda_m}{2 \pi} \left ( 1 + \bm{r}_m
\cdot \bm{\Omega} \right ) ,
\end{equation}
where $\bm{\sigma}$ are the Pauli matrices in the $|j,m
\rangle$ basis, $\sigma_0$ is the identity, $\bm{r}$, and $\bm{r}_m$
are three-dimensional real vectors with $| \bm{r} |, |
\bm{r}_m | \leq 1$, $\bm{\Omega} = ( \sin \theta \cos
\phi , \sin \theta \sin \phi, \cos \theta)$, and $\lambda_m
\geq 0$.

Let us show that for $j=1/2$ no state exhibits nonclassical
properties. This is because for every $\Delta_m$, and using
that $\textrm{tr} \left ( \sigma_j \sigma_k \right ) = 2
\delta_{j,k}$,
\begin{eqnarray}
p_m &=& \textrm{tr} \left ( \rho \Delta_m \right ) = \lambda_m
\left ( 1 + \bm{r}_m \cdot \bm{r} \right ),\nonumber\\
Q_{m,\mathrm{max}} &=& \frac{\lambda_m}{2 \pi} \left ( 1 + |
\bm{r}_m | \right ),
\end{eqnarray}
so that $p_m >p_{b,m}$ would imply $\bm{r}_m \cdot \bm{r} >
| \bm{r}_m |$, which is not possible since $|\bm{r} | \leq 1$.

Similarly, there are no measurements exceeding the classical
bounds since $p_m =\lambda_m ( 1 + \bm{r}_m \cdot
\bm{r} )$, $Q_{\mathrm{max}} = (1+ | \bm{r} |)/(4 \pi)$, and
$\textrm{tr} \Delta_m = 2 \lambda_m$ so that the violation
of the classical bound (\ref{cbsu2a}) would be equivalent
to $\bm{r}_m \cdot \bm{r} > | \bm{r} |$, which is not possible
since $|\bm{r}_m | \leq 1$.

This lack of nonclassical states agrees with the approach in
Ref. \cite{GBB} and with the fact that for $j=1/2$ all pure
states are SU(2) coherent states. On the other hand, this is
in sharp contrast with the fact that all pure state have
negative values of the SU(2) Wigner function \cite{yws}.
Nevertheless, it is worth pointing out that for finite-dimensional
systems the SU(2) distributions such as $P(\Omega)$ and the
Wigner function are nor uniquely defined in contrast with their
infinite-dimensional counterparts \cite{GBB,yws}.

\subsection{$j=1$}

For $j=1$ the coherent states (\ref{cs}) are
\begin{eqnarray}
|1, \Omega \rangle &=& \sin^{2} \left ( \frac{\theta}{2} \right )
|1,-1 \rangle \nonumber \\
&+& \sqrt{2} \sin \left ( \frac{\theta}{2} \right )
\cos \left ( \frac{\theta}{2} \right ) \exp (-i \phi) |1, 0 \rangle
\nonumber \\
&+& \cos^{2} \left ( \frac{\theta}{2} \right ) \exp (-i 2 \phi)
|1,1 \rangle .
\end{eqnarray}

\subsubsection{Nonclassical states}

As a measurement revealing nonclassical states we can consider
the projection on the state $|1, 0 \rangle$ (in the $|j,m
\rangle$ basis),
\begin{equation}
\label{10}
\Delta_0 = \rho_0 = |1, 0 \rangle \langle 1, 0|,
\end{equation}
with $Q$ function,
\begin{equation}
Q_0 (\Omega ) = \frac{3}{8 \pi} \sin^2 \theta,
\quad
Q_{0,\mathrm{max}}= \frac{3}{8 \pi} ,
\end{equation}
so that the classical upper bound in Eq. (\ref{cbsu2b}) is $p_{b,0}
= 1/2$. This is clearly violated  when the measured state is the
same state $\rho = \rho_0$, since in such a case $p_0 = 1 > p_{b,0}
= 1/2$, and there is a 100\% violation of the classical bound.

This agrees with the fact that the state $|1,0 \rangle$ in
Eq. (\ref{10}) can be regarded as the limit of SU(2) squeezed
states \cite{nn1,nn2}. This also agrees with the result in
Ref. \cite{GBB} stating that for $j=1$ classical behavior (i. e.,
nonsingular positive $P(\Omega)$) is equivalent to non-negative
covariance-like matrix
\begin{equation}
Z_{k,\ell} = \langle ( j_k j_\ell + j_\ell j_k ) \rangle -
\delta_{k,\ell} - \langle j_k \rangle \langle j_\ell \rangle .
\end{equation}
For the state $|1,0 \rangle$ in Eq. (\ref{10}) we have \cite{nn2}
\begin{equation}
Z = \pmatrix{1 & 0 & 0 \cr 0 & 1 & 0 \cr 0 & 0 & -1} ,
\end{equation}
so that the state is nonclassical.

\subsubsection{Nonclassical measurements}

We can provide an example of nonclassical measurement. To this end
we consider the measurement (in the $|j,m \rangle$ basis) $\Delta_0
= \rho_0 = |1, 0 \rangle \langle 1, 0|$. As measured state we
consider the phase averaged equatorial SU(2) coherent state
\begin{equation}
\rho = \frac{1}{2 \pi} \int_{2 \pi} d \phi |1, \theta= \pi/2, \phi
\rangle \langle 1, \theta= \pi/2, \phi |,
\end{equation}
where $|j, \theta, \phi \rangle$ are the SU(2) coherent states.
 In the $|j, m \rangle \langle j, m|$ basis this is
\begin{equation}
\rho = \frac{1}{4} |1,1 \rangle \langle 1, 1| + \frac{1}{2} |1,0
\rangle \langle 1, 0| + \frac{1}{4} |1, -1 \rangle \langle 1, -1| ,
\end{equation}
with $Q$ function
\begin{equation}
Q (\Omega ) = \frac{3}{16 \pi} \left ( 1 + \frac{1}{2} \sin^2
\theta \right ),
\quad
Q_\mathrm{max} =  \frac{9}{32 \pi}.
\end{equation}
The probability is $p_0 = 1/2$, which is 167 \% above the
classical bound in Eq. (\ref{cbsu2a}),
\begin{equation}
p_{b} = \frac{4 \pi}{2j+1} Q_\mathrm{max} \textrm{tr} \Delta_0 =
\frac{3}{8}.
\end{equation}

\section{Conclusions}

We have provided feasible practical procedures to reveal the quantum
nature of states and measurements. We have illustrated them with the
most practical measuring schemes available, such as photon-number and
quadrature measurements.

The nonclassical tests proposed in this approach are exceedingly
simple since definite conclusions are obtained without evaluation
of moments, or any other more sophisticated data analysis. This
is reflected on the robustness of these nonclassical criteria under
practical imperfections, such as finite detection efficiencies and
finite sampling.

We have demonstrated that these nonclassical tests are independent of
other typical quantum signatures such as sub-Poissonian statistics,
quadrature squeezing, or oscillatory statistics.

\section*{Acknowledgments}

We thank Dr. Shashank Virmani for fruitful discussions and
Prof. Mark Hillery for valuable comments. A.R. acknowledges financial support from the University
of Hertfordshire and the EU Integrated Project QAP. A.L. acknowledges support from Project No. FIS2008-01267
of the Spanish Direcci\'{o}n General de Investigaci\'{o}n
del Ministerio de Ciencia e Innovaci\'{o}n.

\appendix

\section{Trace measurements}

As we have shown above, the classical bounds for measurements
depend on the trace of the corresponding POVM elements $\Delta_m$.
Incidentally, the trace $\textrm{tr} \Delta_m$ is proportional
to the probability of the outcome $m$ when the input state is
the one of maximum ignorance, $\rho \propto I$, where $I$ is the
identity;
\begin{equation}
p_m (\rho \propto I) \propto \textrm{tr} \Delta_m .
\end{equation}
When $\textrm{tr} \Delta_m$ depends on $m$ this can be regarded
as a kind of prior bias, since some outcomes are more probable
than others even when the measured state presents in principle
no preference for any outcome.

In this appendix we present two simple procedures that allow us
to determine $\textrm{tr} \Delta_m$ in practice. For simplicity
and without loss of generality we focus on the single-mode case.

To this end we note that the identity can be expressed as
\begin{equation}
\label{rI}
I = 2 \int_0^\infty dr r \rho(r),
\end{equation}
where $\rho(r)$ are phase-averaged coherent states (which
correspond to a laser output well above threshold \cite{MSZ})
\begin{equation}
\rho (r) = \frac{1}{2 \pi} \int_{2 \pi} d \phi | r \exp(i\phi)
\rangle \langle r \exp(i\phi) |,
\end{equation}
and  $| r \exp(i\phi) \rangle$ are coherent states $| \alpha
\rangle$ with $\alpha = r \exp(i\phi)$. When illuminating the
detector with phase-averaged coherent states $\rho (r)$ the
measured statistics is essentially the phase average of the $Q$
function of $\Delta_m$,
\begin{equation}
p_m (r) = \frac{1}{2} \int_{2 \pi} d \phi Q_m \left [ r
\exp(i\phi) \right ] .
\end{equation}
After Eq. (\ref{rI}) the desired $\textrm{tr} \Delta_m$ can
be determined by repeating the measurement for different input
states $\rho (r)$ by suitably varying the coherent amplitude
$r$
\begin{equation}
\textrm{tr} \Delta_m = \textrm{tr} \left ( \Delta_m I \right )
= 2 \int_0^\infty r dr p_m (r) .
\end{equation}
Although this recall a tomographic reconstruction of
$\Delta_m$ \cite{MM}, this is not the case since the
illuminating state already carries the angular integration.

The same goal can be achieved following an slightly different
strategy. This is by illuminating the detection system with
thermal-chaotic states (\ref{tc}). The corresponding $Q$ function
(\ref{Qtc}) is a Gaussian centered at the origin of the complex
plane and its width increases when the average mean number of
photons $n_{tc}$ increases. Thus for $n_{tc}$ large enough the
$Q$ function of the thermal-chaotic state will be approximately
constant $Q (\alpha ) \simeq Q_0$ on the area where $P_m (\alpha)
\neq 0$, so that the statistics will be proportional to
$\textrm{tr} \Delta_m$:
\begin{eqnarray}
p_m &=& \pi \int d^2 \alpha P_m (\alpha) Q (\alpha) \nonumber \\
&\simeq& \pi Q_0 \int d^2 \alpha P_m (\alpha) = \pi Q_0 \textrm{tr}
\Delta_m  ,
\end{eqnarray}
and
\begin{equation}
\textrm{tr} \Delta_m  = \lim_{n_{tc} \rightarrow \infty}
\frac{p_m}{\pi Q_0 } .
\end{equation}
To illustrate this idea let us consider a one-photon detector
whose output is contaminated by the vacuum and two-photon
contributions
\begin{equation}
\Delta_1 = q | 0 \rangle \langle 0 |+ p | 1 \rangle \langle 1|
+ q | 2\rangle \langle 2 |,
\end{equation}
with $\textrm{tr} \Delta_1 = p + 2q$. When illuminated with the
thermal-chaotic state we get
\begin{equation}
p_1 = \frac{1}{n_{tc} +1} \left [ q + p \frac{n_{tc}}{n_{tc} +1} +
q \frac{n_{tc}^2}{(n_{tc} +1)^2} \right ].
\end{equation}
For example, for $p = 1$, and $q=0.1$, and taking $\pi Q_0 =
1/(n_{tc}+1)$, we get for $n_{tc} = 100$
\begin{equation}
\textrm{tr} \Delta_1 = 1.2 , \quad
\frac{p_1}{\pi Q_0 } = 1.19 ,
\end{equation}
i. e., only 1 \% error.

\end{document}